\documentclass[preprint]{aastex}

\shorttitle{Testing YSO envelope models with continuum and line 
observations}
\shortauthors{Hogerheijde \& Sandell}
\begin{document}

\title{Testing envelope models of young stellar objects with
submillimeter continuum and molecular-line observations}

\author{Michiel R. Hogerheijde\altaffilmark{1}}
\affil{Radio Astronomy Laboratory, Astronomy Department, University of
California, 601 Campbell Hall, Berkeley, CA 94720-3411}
\altaffiltext{1}{Sterrewacht Leiden, The Netherlands}
\email{michiel@astro.berkeley.edu}
\and
\author{G\"oran Sandell\altaffilmark{2}}
\affil{National Radio Astronomy Observatory, P.O.~Box 2, Greenbank, WV
24944}
\altaffiltext{2}{The National Radio Astronomy Observatory is a
facility of the National Science Foundation operated under cooperative
agreement by Associated Universities, Inc.}

\begin{abstract}
Theoretical models of star formation make predictions about the
density and velocity structure of the envelopes surrounding isolated,
low-mass young stars. This paper tests such models through high
quality submillimeter continuum imaging of four embedded young stellar
objects in Taurus and previously obtained molecular-line
data. Observations carried out with the Submillimeter Continuum
Bolometer Array on the James Clerk Maxwell Telescope at 850 and 450
$\mu$m of L1489~IRS, L1535~IRS, L1527~IRS, and TMC~1 reveal $\sim
2000$ AU elongated structures embedded in extended envelopes. The
density distribution in these envelopes is equally well fit by a
radial power-law of index $p=1.0$--2.0 or with a collapse model such
as that of Shu (1997: ApJ, 214, 488). This inside-out collapse model
predicts $^{13}$CO, C$^{18}$O, HCO$^+$, and H$^{13}$CO$^+$ line
profiles which closely match observed spectra toward three of our four
sources. This shows that the inside-out collapse model offers a good
description of YSO envelopes, but also that reliable constraints on its
parameters require independent measurements of the density and the
velocity structure, e.g., through continuum and line observations.
For the remaining source, L1489~IRS, we find that a model consisting
of a 2000 AU radius, rotating, disk-like structure better describes
the data. Possibly, this source is in transition between the embedded
Class~I and the optically revealed T~Tauri phases. The spectral index
of the dust emissivity decreases from $\beta = 1.5$--2.0 in the
extended envelope to $1.0\pm 0.2$ in the central peaks, indicating
grain growth or high optical depth on small scales. The observations
of L1527~IRS reveal warm ($\gtrsim 30$ K) material outlining, and
presumably heated by, its bipolar outflow. This material comprises
$\lesssim 0.2$ $M_\odot$, comparable to the amount of swept-up CO but
only 10\% of the total envelope mass. Two apparently starless cores
are found at $\sim 10,000$ AU from L1489~IRS and L1535~IRS. They are
cold, 10--15 K, contain 0.5--3.0 $M_\odot$, and have flat density
distributions characterized by a Gaussian of $\sim 10,000$ AU FWHM. The
proximity of these cores shows that star formation in truly isolated
cores is rare even in Taurus.
\end{abstract}

\keywords{stars: formation --- stars: low-mass --- stars: pre-main
sequence --- ISM: dust}

\section{Introduction\label{s:intro}}

Of the two currently identified modes of low-mass star formation,
viz.\ formation of clusters in giant molecular clouds and isolated
star formation in dark molecular clouds, the latter is by far the best
studied \citep*[for recent reviews on star formation, see][]{crete2,
ppiv}. Theoretical models make definite predictions for the density
and velocity structure of these collapsing cores \citep*{shu:selfsim,
terebey:tsc, safier:collapse, mclaughlin:logotropes, foster:collapse},
and several ways exist to test these observationally. Molecular lines
probe the density as well as the velocity distribution, but are
subject to changes in the chemistry and may be affected by the bipolar
outflows driven by many, if not all, young stellar objects
(YSOs). Continuum emission from dust at (sub) millimeter wavelengths
suffers less from these effects, but only traces the column density
integrated along the line of sight and is insensitive to the velocity
field. This paper presents submillimeter continuum observations of
four low-mass YSOs in Taurus: L1489~IRS, L1535~IRS, L1527~IRS, and
TMC~1. We test the often-used model of \citet{shu:selfsim} by using
the model parameters inferred from the continuum emission to predict
line profiles and comparing these to previously obtained spectra.

In the `standard picture' of low-mass star formation, as described,
e.g., in \citet{shu:ppiii}, cloud cores are initially supported by
magnetic fields. As the neutral molecules slip past the field lines
through ambipolar diffusion, the density distribution approaches that
of a singular isothermal sphere: $\rho \propto r^{-2}$
\citep{lizano:ambidiff, basu:ambidiff1}. After its center can no
longer be supported, the core starts to collapse at time $t=0$ from
the inside-out. \citet{shu:selfsim} found a self-similar solution of
the collapse using as spatial coordinate $x=r/r_{\rm CEW}$, where
$r_{\rm CEW}=a\,t$ is the location of the head of the collapse
expansion wave with $a$ the sound speed. Just inside $r_{\rm CEW}$,
the density profile is relatively flat, with $\rho \propto r^{-1}$,
but steepens to $\rho \propto r^{-1.5}$ as material approaches
free-fall further in. In other words, the density profile can be
roughly represented by $\rho \propto r^{-p}$ with $p=1$--2. The
corresponding velocities increase from $v=0$ to $v\propto r^{-0.5}$,
characteristic of free fall.

The inside-out collapse model constructed by \citet{shu:selfsim} is a
very simple one, and many authors have extended it to larger degrees
of realism, including rotation \citep{terebey:tsc} and magnetic fields
\citep{galli:pseudo1, li:toroids, li:isopedic, galli:isopedic}. The
directional dependence of cloud support in these cases leads to the
formation of a flattened structure at the center of the collapsing
envelope, merging into a centrifugally supported circumstellar
disk. Entirely different approaches to cloud collapse are proposed by
\citet{whitworth:selfsim, foster:collapse, mclaughlin:logotropes}; and
\citet{safier:collapse}. However, the simplicity of
\citeauthor{shu:selfsim}'s model has made it very popular for the
analysis of continuum and line data. Many authors have confirmed its
success in explaining the broad-band spectral energy distributions
(SEDs) of embedded YSOs \citep*{adams:ysosed, ladd:firsubmm1}. Their
envelopes follow a radial power law with negative index between 1 and
2 \citep*{ladd:firsubmm2, chandler:envelopes, mrh:serpens}, while cores
which do not (yet) contain an embedded star have density distributions
which are much flatter at their center
\citep*{wardthompson:preprotostellar, andre:l1698b, motte:rhooph}.
Molecular-line profiles observed toward YSO envelopes can be fitted
with the inside-out collapse model as well
\citep*[e.g.][]{zhou:b335collapse, zhou:rotation, zhou:l1527+cb54,
choi:b335mc, saito:b335, mrh:thesis}. The infalling motions of
collapse models result in the so-called infall asymmetry in
self-absorbed lines profiles, with the blue peak stronger than the red
one. This signature is detected toward a majority of deeply embedded
objects \citep{mardones:infall, gregersen:infall}.

This paper investigates collapse models as descriptions of the
envelopes around embedded YSOs through 850 and 450 $\mu$m observations
of $0.15\times 0.15$ pc fields with the powerful Submillimeter Common
User Bolometer Array (SCUBA) on the James Clerk Maxwell Telescope
(JCMT). We use the inside-out collapse model of \citet{shu:selfsim} as
a template for collapse models because of its simplicity and because
it is so often invoked in the literature: a critical investigation of
this model's ability to fit the high quality continuum data from SCUBA
together with molecular line profiles seems warranted.  The four
sources of our sample (L1489~IRS = IRAS 04016+2610; L1535~IRS =
04325+2402; L1527~IRS = 04368+2557; and TMC~1 = 04381+2540) are all
low-mass embedded Class~I YSOs \citep*[see][for definitions of YSO
classes]{lada:iau115,andre:vla1623}; L1527~IRS particularly deeply
embedded and sometimes referred to as a Class~0 object. They have low
bolometric luminosities of 0.7--3.7 $L_\odot$ and estimated stellar
masses $<0.4$ $M_\odot$ \citep[for a more detailed description of
their characteristics, see][]{mrh:taurus1, mrh:taurus2}. Previous
continuum observations, of lower signal-to-noise and covering smaller
fields of view, are reported by \citet{ladd:firsubmm1, ladd:firsubmm2}
and \citet{chandler:envelopes}. Recently, \citet{chandler:scuba} and
\citet{shirley:scuba_baas} reported very similar results using SCUBA
observations of YSOs including L1527~IRS. Our work stands apart from
these papers in that it also considers the velocity structure of the
envelopes as measured through molecular lines.

After describing the observations and their reduction (\S
\ref{s:obs}), section \ref{s:results} discusses the characteristics of
the emission. Section \ref{s:models} first analyzes the extended
emission in terms of a power-law density distribution (\S
\ref{s:powmdl}) followed by the inside-out collapse model (\S
\ref{s:shumdl}). We then verify the derived model parameters by
comparing the calculated line profiles to observed spectra (\S
\ref{s:molmdl}). The paper continues with a discussion of the
implications of our modeling (\S \ref{s:tests}) and of the properties
of L1489~IRS (\S \ref{s:l1489}), for which a different model
description seems required. Section \ref{s:central} discusses in more
detail the central 2000 AU of the envelopes, where different
conditions may exist. The properties of two starless cores identified
near L1489~IRS and L1535~IRS are discussed in \S \ref{s:coreb}. A
brief summary of our main findings concludes the paper in section
\ref{s:conclusions}.

\section{Observations and reduction\label{s:obs}}

We observed the submillimeter-continuum emission of four low-mass YSOs
in Taurus in 1997 September and December, and 1998 January using SCUBA
on the JCMT\footnote{The James Clerk Maxwell Telescope is operated by
the Joint Astronomy Centre on behalf of the Particle Physics and
Astronomy Research Council of the United Kingdom, the Netherlands
Organization for Scientific Research and the National Research Council
of Canada.}, on Mauna Kea, Hawaii. SCUBA is a bolometer array camera
with a long-wavelength array of 37 pixels and a short-wavelength
array of 92 pixels. Both arrays can be used simultaneously.  In our
observations the arrays were set to 850 and 450 $\mu$m,
respectively. \citet{holland:scuba} describes the instrument and its
observing modes in detail; this section only gives a short
description of our observations and their specific details.

All observations were done as 64-point `jiggle' maps, where the
secondary mirror is used to produce a Nyquist-sampled image of the 450
$\mu$m array by chopping at a frequency of 7.8125 Hz. To cancel sky
variations this jiggle pattern is split into four 16-point
sub-patterns of 1 second integration time per point, after which the
telescope nods to the negative beam and repeats the pattern. The
complete 64-point pattern takes 128 seconds, with 64 seconds spent
looking at the sky on each side of the array. Because of this
sky-subtraction, jiggle-maps can only be made of sources which are
smaller along the chop direction than the throw distance. In practice
this translates to $\sim 120''$; larger chop throws result in poor sky
subtraction and degradation of the image quality.

Most of our observations were done in service mode and always in
stable night-time conditions with an initial $120''$ chop in
azimuth. If the observations spanned several nights, we inspected the
first night's results and modified the chop throw or the offset of the
array to include as much extended emission as possible. Each observing
sequence started with a pointing check on a nearby blazar or secondary
calibrator, followed by 5--10 integrations on the target and
another pointing check. Any pointing drifts were corrected for by
linear interpolation in azimuth and elevation.

Table \ref{t:obs} lists the best-fit source positions, observing
dates, number of integrations, and chop throw position angle. Most
observations were done in excellent or good sub-millimeter conditions,
with typical zenith opacities at 850 $\mu$m of 0.1--0.3 and seeing of
0{\farcs}2--0{\farcs}5.  Only a few observations were done during
marginal sub-millimeter nights, in which case we only used the 850
$\mu$m data.  In 1997 September we used Uranus as a primary
calibrator, but for our observations in 1997 December and 1998 January
we only used the secondary calibrators CRL~618 and HL~Tau. The 850
$\mu$m and 450 $\mu$m optical depth was estimated from the 1.3~mm
opacity monitor at the Caltech Submillimeter Observatory, and
occasionally double-checked with 850 $\mu$m sky dips with SCUBA.

The data were reduced in a standard way using the SCUBA reduction
package SURF \citep{jenness:surf}. We took extreme care to inspect the
images and blank out any bolometers suspected to have chopped onto
source emission. This could not be avoided for L1535~IRS, since the
data were taken in only one night and the chop direction could not be
corrected later. The images of L1489~IRS are also likely to be
affected at their northeast edge, where the emission goes to zero
coincident with the chop direction. The cloud emission also drops off
fast in the southwest, but an observation with $150''$ chop throw as
well as previous HCO$^+$ 1--0 and C$^{18}$O 2--1 data suggest that
this morphology is real. In all cases, the drop off of emission beyond
$120''$ is due to the chop throw; this only affects the emission near
the edges of the images, but not on smaller scales, as confirmed from
an observation of L1527~IRS with a $90''$ throw.

After calibration, pointing correction, and sky noise subtraction of
each scan, the 850 $\mu$m images were coadded to determine the
best-fit positions of the submillimeter source in each field (Table
\ref{t:obs}). We then repeated the coadding, now including proper
weighting with the noise levels and aligning each scan to the best-fit
850 $\mu$m position. We used the 850 $\mu$m position to align the data
from the 450 $\mu$m array, which is slightly offset from the
long-wavelength array. This shift and add procedure ensures that the
final images are as sharp as possible. The SCUBA positions agree to
within $1''$ of available aperture-synthesis values for L1489~IRS and
L1527~IRS \citep{rodriguez:radiocores, mrh:taurus1}. The latter work
only marginally detected L1535~IRS and TMC~1, and SCUBA significantly
improves their known positions. For TMC~1 we find offsets of
$(\Delta\alpha,\, \Delta\delta) = (+4{\farcs}5,\, -2{\farcs}7)$ from
the position given there, which should be accurate to within
$1''$. The observations of L1535~IRS suffer from unusually large,
systematic pointing drifts, but an offset of $(+4{\farcs}5, \,
+5{\farcs}4)$ is obtained with an uncertainty of $2''$ in right
ascension and $1''$ in declination estimated from map to map scatter.

The images were further analyzed with the MIRIAD software package
\citep*{sault:miriad}. Because of the shape of the JCMT's primary
reflector, deviations of its beam from a single Gaussian are an
important factor in the analysis. At short wavelengths, for example,
equal power lies within the $8''$ FWHM main beam as in a much broader
error beam.  Unfortunately, the lack of observable planets during most
of our observations precluded the contemporaneous measurement of the
beam profile. We have used observations of Uranus from 1997 September
to construct an ad-hoc model of the beam. Fitting the azimuthally
averaged Uranus data, we find a satisfactory description of the beam
profile by three Gaussians with relative amplitudes and FWHM as listed
in Table \ref{t:beam}. The $120''$ dimension of the largest beam
component is partially determined by the size of the chop throw, and
thus reflects the beam pattern set by the JCMT dish and the observing
mode. This description of the beam is used in the analysis below.

\section{The submillimeter-continuum images\label{s:results}}

The 850 and 450 $\mu$m images show emission peaks at the position of
the young stars, surrounded by extended emission filling almost the
entire fields of view (Fig. \ref{f:maps}). Second emission peaks are
visible $\sim 1'$ northeast of L1489~IRS and north of L1535~IRS.
Table \ref{t:flux} lists the total flux contained in the images, and
results of a Gaussian fit to the central emission peaks. These fits
reflect the smallest sized Gaussians which can describe the emission
peaks, while treating the extended emission as an unrelated background
(see \S \ref{s:central}).  The fluxes and morphologies are consistent
with previously published results \citep{ladd:firsubmm2,
moriarty:yso2, chandler:envelopes, chandler:scuba}. A blow-up of the
$40''\times 40''$ around the objects at 450 $\mu$m
(Fig. \ref{f:central}) show elongated cores oriented roughly
perpendicular to the known outflow directions. The fit results of
Table \ref{t:flux} yield aspects ratios between 3:2 and 2:1 for the
central regions of L1489~IRS, L1527~IRS, and TMC~1 after deconvolution
of the beam size. The morphology of L1535~IRS's emission peak is not
well defined and a correspondingly larger size of $12''\times 9''$ is
found. Its outflow direction is also poorly known. The position angle
of $10^\circ$ quoted by \citet{mrh:taurus2} was based on scattered 2
$\mu$m emission interpreted as an outflow cavity, where in fact it is
light scattered off the second submillimeter core north of L1535~IRS
(see \S \ref{s:coreb}). The elongated core of L1489~IRS and the
irregular structure of L1535~IRS's emission peak closely resemble
aperture-synthesis maps of HCO$^+$ and $^{13}$CO $J$=1--0
\citep{mrh:taurus2}.

When plotted on a log-log scale in Fig. \ref{f:azim}, which better
brings out the extended emission, the azimuthally averaged, radial
emission profiles of the sources indicate that the extended emission
follows a radial power law, corresponding to a straight line in the
figure. For L1527~IRS only the radial profile perpendicular to the
east-west outflow is shown. L1489~IRS appears much more compact than
the other sources, a result which does not depend on the likely
artificial drop in emission at the northeast edge of the images (\S
\ref{s:obs}).  In addition to emission directly associated with the
YSOs, L1489~IRS and L1535~IRS show second emission peaks at
$(+60'',+27'')$ and $(+31'',+61'')$, respectively, or $\sim 9500$ AU
from the YSOs. While L1489~IRS's observations were recentered to cover
the emission of this second core, the peak adjacent to L1535~IRS is
located near the edge of the images. Since it dominates the emission
over the entire field of view, it severely hampers the analysis of the
L1535~IRS data. The emission around L1527~IRS is elongated along the
east-west outflow, perpendicular to the orientation of its central
peak. The outflows of the other sources, which have kinetic
luminosities smaller by factors of 3--100 \citep{mrh:taurus2}, do not
leave a detectable imprint on the dust emission.

The quality of the SCUBA observations is sufficient to derive the
spectral index $\alpha$ between 850 and 450 $\mu$m (Figs. \ref{f:maps}
and \ref{f:azim}).  To obtain images of $\alpha$, we deconvolved the
850 and 450 $\mu$m images with the appropriate beam profiles,
convolved them with a single Gaussian to the same resolution of
14{\farcs}5, and calculated $\alpha = \log(F_{450}/F_{850}) / \log(850
\,\mu{\rm m} / 450\,\mu{\rm m})$. The resulting values of $\alpha$
range between 1.8 and 3.5. The largest values are found toward the
position of L1489~IRS and along the outflow cavity of L1527~IRS; the
lowest values around TMC~1.  A proper description of the beam profile
is crucial when deriving the spectral index: using a single Gaussian
for the beam lowers $\alpha$ by 0.5 on average and significantly
changes its spatial distribution.

In the submillimeter range, the overall spectral index $\alpha$
reflects the emission-averaged values of the dust temperature, the
opacity, and the spectral index of the dust emissivity
$\beta$. Assuming fully optically thin emission and $\beta=$1.5--2.0
\citep*[as expected for dust in dense clouds,][]{goldsmith:codust,
ossenkopf:kappa}, the observed $\alpha$ directly translates to dust
temperatures of 10--15 K for the extended material and $\gtrsim 30$ K
for the material near L1489~IRS and along the outflow of L1527~IRS;
lower values of $\beta$, corresponding to grain growth
\citep{ossenkopf:kappa,pollack:kappa}, give higher temperatures. At
the position of L1527~IRS, the spectral index shows a local minimum of
$\sim 2.5$. The next section will show that rather than a lower
temperature, a decreased $\beta$ in the central region or a
significant contribution from optically thick emission from an
unresolved circumstellar disk are more likely explanations. The
extended emission peaks northeast of L1489~IRS and north of L1535~IRS
do not show up as extrema in $\alpha$, indicating that they correspond
to enhancements in column density rather than temperature. In the
following, we will refer to these condensations as L1489~NE-SMM and
L1535~N-SMM.

\section{Modeling the structure of the YSO envelopes \label{s:models}}

\subsection{Power-law density models\label{s:powmdl}}

The emission profiles of Fig. \ref{f:azim} show that the emission of
the extended emission follows a radial power law. The inside-out
collapse model predicts a power-law distribution for the density with
negative index between 1 and 2, where the exact shape is given by the
location of the collapse expansion wave $r_{\rm CEW}$. In this section
we investigate which single value of the power-law index $p$ best fits
the overall emission around the sources; the next section will focus
on the inside-out collapse model. Since these model results are mostly
constrained by the extended emission, no additional unresolved sources
like circumstellar disks are included \citep[but see][for a discussion
of the effect of the presence of an unresolved central source on
derived model parameters]{chandler:scuba}. These models are
spherically symmetric; at the end of this section (\S \ref{s:central})
we will look at the properties of the central elongated emission
peaks, and how they differ from those inferred for the envelope as a
whole.

For the power-law model we adopt a density distribution $\rho = \rho_0
(r/1000\,{\rm AU})^{-p}$, where $\rho_0$ is the density at an
arbitrary radius of 1000 AU. Instead of density $\rho$ we will use the
H$_2$ number density $n$ as a parameter, assuming a `standard'
gas-to-dust ratio of 100:1. The inner radius of the envelope is set at
50 AU, which does not influence the results, while the outer radius is
initially set at 8000 AU. Following \citet{adams:irsed}, we assume the
dust temperature to follow a radial power law with index 0.4, $T_{\rm
d} \approx 26 (r/1000\,{\rm AU})^{-0.4} (L_{\rm
bol}/1\,L_\odot)^{0.2}$ K. Such a distribution is expected for a
spherical cloud with an embedded heating source and which is optically
thin to the bulk of the heating radiation. The dust temperature
depends on the luminosity of the central source as described by
\citet{adams:irsed}. For our sources, $T_{\rm d}(1000\,{\rm AU})=34$ K
(L1489~IRS: 3.7 $L_\odot$) and 24 K (L1535~IRS and TMC~1: 0.7
$L_\odot$). These temperature distributions reproduce the observed
SEDs at millimeter and infrared wavelengths for the envelope
parameters derived below \citep*[e.g.,][]{ladd:firsubmm1,
kenyon:sed_1}. For L1527~IRS, the above expression yields $T_{\rm
d}(1000\,{\rm AU})=27$ K, which significantly overestimates the IRAS
fluxes at 100 and 60 $\mu$m. For the parameters derived below, the
opacity of its envelope is around unity at these wavelengths, and it
is possible that a flattened rather than spherical geometry of its
edge-on envelope could increase the opacity by a factor of a few and
hence reconcile the fluxes. However, for our modeling we choose to
lower the temperature of L1527~IRS's envelope to $T_{\rm d}(1000\,{\rm
AU})=18$ K, in which case the 100 and 60 $\mu$m are reproduced.

The power-law index of $q=0.4$ of our adopted temperature profile
($T_{\rm d} \propto r^{-q}$) is strictly true only for
$\beta=1$. \citet{chandler:scuba} give an alternative description of
$T_{\rm d}(r)$, where $q=2/(4+\beta)$. For realistic values of $\beta$
of 1--2, this translates to $q=0.33$--0.4. Heating by the external
radiation field may be an other factor influencing the temperature
profile \citep[e.g., see][]{choi:b335mc}. Since the Rayleigh-Jeans
limit of the Planck function is not valid in the submillimeter range,
the emission -- and the inferred model parameters -- depend
sensitively on the temperature distribution. Provided that the choice
for the adopted temperature distribution is reasonable, the
conclusions that a certain model does or does not fit the data is more
robust that the exact value of the inferred model parameters. Since it
is the main aim of this paper to investigate models for YSO envelopes,
rather than derive accurate constraints on their parameters, we deem
our simple description of the temperature distribution sufficient, but
urge the reader to keep in mind that the values of the inferred
parameters are valid only within the framework of our model
assumptions.

Realistic models for the dust emissivity at submillimeter wavelengths,
including grain growth, are available \citep{ossenkopf:kappa,
pollack:kappa}, but for the sake of simplicity we parameterize the
dust emissivity as $\kappa = \kappa_0 (\nu/\nu_0)^\beta$, with
$\nu_0=10^{12}$ Hz and $\kappa_0 = 0.1$ cm$^{2}$~g$^{-1}$ (gas and
dust, with a gas:dust ratio of 100:1; see
\citealt{hildebrand:kappa}). Typical values for $\beta$ range between
1 and 2, although much lower values of $\beta \approx 0$ have been
found in T~Tauri disks \citep{beckwith:beta}. The absolute values for
$\kappa$ lie within a factor of 2--3 from those of the more elaborate
models, a difference which only influences the inferred mass, not the
density power law index.

The free parameters of this model are the density at 1000 AU $n_0$,
the density power-law index $p$, and the dust spectral index $\beta$.
These correspond to the observed total flux, radial emission profile,
and ratio of 450 to 850 micron emission, respectively.  For each
combination of $(n_0, p, \beta)$ we calculate the emission from the
envelope model over the imaged region by following the radiative
transfer along a large number of lines of sight through the source
using the Planck function. The resulting intensity distribution is
then convolved with the appropriate beam pattern, and a $\chi^2$
measure is determined of the difference between the model and the data
at both wavelength bands together over a $80''\times 80''$
($11,200\times 11,200$ AU) region around the source. Minimization of
the $\chi^2 \equiv \Sigma (F({\rm model})-F({\rm obs}))^2/\sigma^2$,
where $F$ is the intensity in Jy~beam$^{-1}$, $\sigma$ the noise, and
the summation is over all pixels within the $80''\times 80''$ region
and both wavelength bands, yields best-fit values of $n_0$, $p$, and
$\beta$ listed in Table \ref{t:powfit}. The indicated uncertainties in
the best-fit values include the effects of signal-to-noise, the shape
of the $\chi^2$ surface, and an estimated 20\% calibration uncertainty
in the 850 and 450 $\mu$m data. The reduced $\chi^2$ values of these
fits are 1.5 for L1489~IRS, 0.8 for L1535~IRS, 2.7 for L1527~IRS, and
0.7 for TMC~1.

The model fit requires assumptions about the contribution of
L1489~NE-SMM and L1535~N-SMM to the emission; the emission along
L1527~IRS's outflow does not contribute significantly within the
fitting region. We assume a Gaussian emission distribution centered on
the former cores with adjustable flux at 850 $\mu$m, spectral index,
and major and minor axes. This increases the number of free parameters
from three to seven. Because the $\chi^2$ measure is still limited to
$80''$ around the YSOs, the best-fit values for the latter four
parameters are not representative for the L1489~NE-SMM and L1535~N-SMM
cores as a whole but only describe the `background' emission around
the source due to the secondary cores (see \S \ref{s:coreb}).

The observations are well fit by the power-law model (Fig.
\ref{f:azim}), if the outer radius is decreased to 2000 AU for
L1489~IRS and increased to 12,000 AU for TMC~1. This is required to
fit the compact size of L1489~IRS and the emission out to the edge of
TMC~1's images. The radii of 8000 and 12,000 AU reflect the imaged
regions, not necessarily the true extent of the sources. The derived
values of $n_0$, $p$, and $\beta$ are constrained by the central
regions of the envelopes and do not depend critically on $R_{\rm
out}$; only for L1489~IRS would $p$ increase from 2.1 to 2.5 if
$R_{\rm out}=8000$ AU were used. L1527~IRS is well fit with $R_{\rm
out}=8000$ AU, while the dominance of L1535~N-SMM precludes strong
constraints on the size of L1535~IRS itself.  For our sources, the
inferred power-law indices $p$ range between 0.9 for L1527~IRS and 2.1
for L1489~IRS, and envelope masses are 0.016--3.7 $M_\odot$. For
L1527~IRS \citet{chandler:scuba} infer an equally low index $p$ of
1.0--1.2 from similar SCUBA data. For TMC~1 our inferred value of
$p=1.2 \pm 0.1$ is marginally consistent with the result of SED
fitting by \citet{chandler:envelopes} of $p=0.9\pm 0.3$.  The
difference is possibly due to the much smaller outer radius of 2500 AU
adopted by these authors corresponding to their smaller imaged
regions. The inferred values for the spectral index $\beta$ of the
dust emissivity are 1.5--2.0, excluding the uncertain values for
L1535~IRS, as expected for material in dense molecular clouds which
underwent some grain growth \citep[$\beta \approx
1.5$,][]{ossenkopf:kappa, pollack:kappa}.

\subsection{The inside-out collapse model\label{s:shumdl}}

The data are well described by a radial power-law distribution with
$p=1$--2.  This lies in the range of indices predicted by the
inside-out collapse model. Can this collapse model, which has a slope
varying with radius, also describe the data? In this section we fit
the inside-out collapse model of \citet{shu:selfsim} to the data,
where we parameterize the model in $a$ and the location of the
collapse expansion wave, $r_{\rm CEW}=a\,t$. This parameterization
offers a good match to the directly observed quantities peak flux
($\Leftrightarrow a$) and shape of the emission distribution
($\Leftrightarrow r_{\rm CEW}$). The other model parameters
(temperature, inner and outer radius, dust emissivity) are the same as
in the previous section. We again adopt a Gaussian model for
L1489~NE-SMM and L1535~N-SMM, without which no satisfactory model fit
can be found, which increases the number of free parameters for these
objects from three to seven. Best-fit parameters follow from $\chi^2$
minimization over $80''\times 80''$ regions around the YSOs (Table
\ref{t:shufit}), yielding reduced $\chi^2$ values of 1.5 for
L1489~IRS, 0.7 for L1535~IRS, 3.0 for L1527~IRS, and 0.8 for
TMC~1. The table does not list the parameters of the Gaussian model
for L1489~SE-SMM and L1535~N-SMM, which are essentially similar to
those found in \S \ref{s:powmdl} and listed in Table \ref{t:powfit}.
  
The inside-out collapse model does not predict a density distribution
characterized by a single power-law index throughout the
envelope. However, the beam sizes of $8''$--$14{\farcs}5$ (1000--2000
AU) are sufficiently large compared to the envelopes, that the
resulting distribution of the emission after beam convolution is not
far off from a single power law. The observations can therefore be
equally well fit with the inside-out collapse model as with single
power laws, as witnessed by the very similar reduced $\chi^2$ values
and model curves in Fig.  \ref{f:azim}. The inferred masses and
spectral indices of the dust emissivity are close to those found from
the power-law fitting; only those of L1535~IRS differ somewhat,
probably because of the larger uncertainty in its derived parameters
due to the dominance of L1535~N-SMM.

For L1527~IRS and TMC~1 sound speeds and ages are inferred of 0.44 and
0.19 km~s$^{-1}$, and $3\times 10^4$ and $3\times 10^5$ yr,
respectively. Again, an outer radius of 12,000 AU is used for TMC~1,
which does not influence the values of $a$ or $r_{\rm CEW}$. For
L1535~IRS more uncertain values of 0.29 km~s$^{-1}$ and $6\times 10^5$
yr are found. L1489~IRS can only be fit satisfactory for an outer
radius of 2000 AU, because larger radii require a density fall-off
which is too steep for the model ($p\approx 2.5$, \S
\ref{s:powmdl}). For $R_{\rm out}=2000$ AU, we find a high sound speed
of 0.46 km~s$^{-1}$ and large age of $2\times 10^6$ yr. This
corresponds to a location for the collapse expansion wave of $2\times
10^5$ AU, much larger than $R_{\rm out}$. For the other sources,
$r_{\rm CEW}$ is slightly smaller than $R_{\rm out}$, and L1489~IRS's
value suggests that a different model may be required for this source.

\subsection{Probing the velocity field with molecular lines\label{s:molmdl}}

The previous two sections have shown that the continuum observations
are well fit by a density distribution following a radial power-law of
index $p=1$--2. Many collapse model predict distributions with slopes
in this range, and the inside-out collapse model of
\citet{shu:selfsim} is therefore found to satisfactorily represent the
data. However, a collapse model implies infalling motions. Several
authors have used molecular-line observations to derive -- very
similar -- model parameters for our sources \citep{zhou:l1527+cb54,
mrh:thesis}. The results obtained by \citet{mrh:thesis}, listed in
Table \ref{t:oldshu}, closely agree for TMC~1 with those inferred
above from the continuum observations, but not in the case of the
other sources. For L1527~IRS the value for $r_{\rm CEW}$ is similar,
but its sound speed is lower and its age is correspondingly larger by
a factor of two. Even larger differences are found for L1489~IRS and
L1535~IRS. Rather than comparing inferred model parameters, which may
suffer from systematic effects because of different model assumptions,
a more direct test is to look at the line profiles predicted by the
inside-out collapse model and the best-fit parameters of Table
\ref{t:shufit}. How well are the observed spectra reproduced by the
profiles calculated using these parameters derived directly from the
SCUBA continuum observations?

Using single-dish observations of $^{12}$CO, $^{13}$CO, C$^{18}$O,
HCO$^+$ and H$^{13}$CO (previously presented by \citealt{mrh:taurus1,
mrh:taurus2}, and shown in Fig. \ref{f:spec}), we calculate line
profiles based on the same parameters as derived from the continuum
emission (Table \ref{t:shufit}). The modeling of the molecular
excitation and line radiative transfer employs a spherically-symmetric
Monte-Carlo method developed by \citet{mrh:code}. The envelopes are
divided into 32 concentric shells, sufficient to follow the excitation
and optical depth. The maximum optical depth encountered in the model
is $\approx 30$, and 32 shells ensure that all shells are optically
thin. A calculation with 64 shells confirmed that the results are
independent of the adopted gridding. We adopt a `standard' CO
abundance of $10^{-4}$ with respect to H$_2$, and isotopic ratios of
$^{12}$C:$^{13}$C of 65:1 and $^{16}$O:$^{18}$O of 500:1
\citep{wilson:abundances}. The optically thin H$^{13}$CO$^+$ lines
constrain the HCO$^+$ abundance to $\sim 1\times 10^{-9}$ for
L1527~IRS, $\sim 2\times 10^{-9}$ for L1489~IRS, $4\times 10^{-9}$ for
TMC~1, and $7\times 10^{-9}$ for L1535~IRS. Initially, we assume that
the kinetic temperature is equal to the dust temperature throughout
the envelope. The adopted local turbulent line width of 0.2
km~s$^{-1}$, independent of radius, is similar to that found in dark
cloud cores \citep{fuller:linewidths, myers:tnt}; since the systematic
(infall) motions of the model dominate the overall velocity field, the
calculated profiles are independent to changes in the turbulent width
of factors of a few. After the level populations have converged, the
sky brightness distribution is calculated and convolved with the
appropriate beams.

With these parameters, the modeled line profiles are wider and more
intense than observed. However, for L1527~IRS and TMC~1 much better
agreement between the models and the observations if found
(Fig. \ref{f:spec}) when CO is depleted by a factor of 30 in regions
with $T_{\rm kin}<20$ K \citep[the sublimation temperature of
CO,][]{sandford:sublime_co_co2_h2o}. In the cold and dense envelopes
around YSOs, freezing out of CO on dust grains is responsible for
observed depletions by factors of 10--20 \citep[e.g.,][]{gab:n1333i4,
mundy:iau178, kramer:ic5146}. Obvious exceptions where the models do
not reproduce the $^{12}$CO observations are self-absorption due to
low-density foreground material and contributions from the outflows to
the line wings. The HCO$^+$ line profiles are particularly well fit,
supporting the conclusion of \citet{mrh:taurus1} that this species is
an excellent tracer of YSO envelopes. The calculations do not include
any depletion for HCO$^+$, but use the abundance which is directly
constrained from the H$^{13}$CO$^+$ data.

For L1535~IRS the predicted $^{12}$CO and HCO$^+$ lines are still
slightly wider than observed, even with CO depletion. The model
predicts much lower intensities for the optically thin lines of
$^{13}$CO and C$^{18}$O than observed. This suggests that these lines
primarily trace the low density gas associated with L1535~N-SMM. Since
the mismatch in $^{12}$CO and HCO$^+$ is most likely due to the
uncertain nature of the derived parameters because of L1535~IRS's weak
continuum emission, trying to obtain a closer fit is probably not a
very useful exercise. For L1489~IRS, the line profiles predicted by
the model calculations do not agree at all with the observations,
showing much larger widths (10--15 km~s$^{-1}$ vs. $\sim 4$
km~s$^{-1}$). Apparently, its compact size and steep density fall-off
are not well described by the inside-out collapse model (see \S
\ref{s:l1489}).

To answer the question posed at the beginning of this section: yes,
the inside-out collapse model can reproduce the observed line
profiles.  And where it fails, it confirms earlier suspicions that the
inside-out collapse model is not a good description for L1489~IRS.

\section{Discussion\label{s:discussion}}

\subsection{Critical tests of collapse models\label{s:tests}}

Section \ref{s:shumdl} showed that the inside-out collapse model can
successfully describe the density and the velocity structure of the
extended envelopes around our YSOs. Perhaps the most direct indication
that the line and continuum data are tracing the same collapse model,
is given by the HCO$^+$ line widths toward L1527~IRS and TMC~1. The
continuum data suggest a factor of two difference in sound speed,
which is, as the inside-out model predicts, reflected in the factor of
two difference in line widths seen toward the sources
(Fig. \ref{f:spec}; note the different respective horizontal
scales). Our modeling clearly illustrates that a full test of a
collapse model requires independent measurements of the density and
the velocity structure. We have used submillimeter continuum and line
observations to obtain this goal. A judicially chosen set of molecules
and lines will work as well, if some of these lines are readily
thermalized and their excitation no longer depend on the density.

Combining continuum and line data, as well as combining lines from
different molecules, introduces some level of uncertainty, however. In
our case, it arises from uncertainties in the dust emissivity, the CO
depletion, and possibly the gas-to-dust ratio. For a molecular line
set it is reflected in the abundances of the different species. In
general, it comes down to the question of how the fluxes in different
tracers are scaled to one another. Hence, it is easier to test whether
a certain collapse model can fit the data than to accurately constrain
its parameters. In \S \ref{s:shumdl} we confirmed that the envelopes
of L1527~IRS and TMC~1 are well described by the infall model, but the
derived values for the sound speed and the age are valid only for the
adopted values for the dust emissivity and the dust temperature
distribution. For example, different values are quoted in \S \ref{s:molmdl}
derived from directly fitting line profiles, a process which comes with
its own set of model assumptions (kinetic temperature, abundance, etc.).

Clearly, this situation can be much improved if the temperature
distribution of the dust and of the gas is constrained by
observations. Self-consistent modeling of the continuum radiative
transfer and the heating and cooling balance of the dust and the gas
will be an important step, as will be observations of spatially
resolved SEDs (e.g., from SOFIA) and sensitive tracers of the gas
kinetic temperature (e.g., H$_2$CO, \citealt{mangum:h2co}). With these
tools, we can hope to go beyond the simple inside-out collapse model
of \citet{shu:selfsim} and investigate more realistic, but more
complex, collapse models.

\subsection{A different model for L1489~IRS\label{s:l1489}}

For L1489~IRS the inside-out model spectacularly fails to
simultaneously fit the continuum and line data, although both data
sets can be fit separately (see \S \ref{s:shumdl} and
\citealt{mrh:thesis}). Together with other pieces of evidence, this
leads us to suggest that L1489~IRS is not embedded in a collapsing
envelope, but is instead surrounded by a 2000 AU radius, rotating,
thick disk-like structure. First, Fig. \ref{f:azim} indicates a much
more compact size of $\sim 2000$ AU in radius compared to the other
sources. Second, interferometric observations of HCO$^+$ and $^{13}$CO
1--0 directly reveal a rotating structure of this size, coincident in
orientation with the position angle of Table \ref{t:flux}
\citep{mrh:taurus2}. Third, its comparatively bright near-infrared SED
appears intermediate between that of embedded objects and T~Tauri
stars \citep{ladd:firsubmm1}. In such a disk-like structure, one still
expects power-law distributions for the density and the velocity
(Keplerian rotation), but a different distribution of mass with
velocity. Therefore, the power-laws of the inside-out collapse model
can still fit the density and the velocity individually but not
simultaneously. Our data do not constrain the vertical structure of
this disk-like structure, since it is almost unresolved along its
minor axis.

Could L1489~IRS be a transitional object between the fully embedded
Class~I and the optically revealed T~Tauri phases, with its extended
envelope collapsed into a 2000 AU radius disk-like structure? Its
inclination has been constrained to near edge-on
($i=60^\circ$--$90^\circ$; \citealt{kenyon:sed_2}), and a thick
disk-like structure may explain its bright near-infrared but visually
obscured nature \citep[see also][for HST/NICMOS observations of this
same structure]{padgett:nicmos}. This situation is reminiscent in some
ways of T~Tau, where a Class~II star (T~Tau~N) also has many of the
characteristics of a Class~I object because of the large amount of
circumstellar material -- in fact, T~Tau consists of a Class~II object
T~Tau~N and a more embedded, Class~I companion, T~Tau~S
\citep*{dyck:ttaus}. That T~Tau~N is classified as an optically
visible Class~II object may be largely due to its favorable face-on
orientation with respect to the observer \citep*{herbst:ttauwind},
possibly exactly opposite from the Class~I classification of the more
edge-on L1489~IRS. Other objects which come to mind in this context
are HL~Tau \citep[see, e.g.,][]{stapelfeldt:hltau}, which also has a
steep density distribution in its envelope \citep{chandler:scuba}.

An important difference with T~Tau, and with many younger Class~I
objects, is that no (100 AU radius) centrifugal disk of significant
mass appears to surround L1489~IRS. It is only marginally detected in
3~mm aperture-synthesis observations which selectively trace such
disks \citep*[see][]{terebey:disks}. It seems unlikely that L1489~IRS
is a low-mass system, and hence may have an intrinsically low-mass
disk, because of its comparatively large luminosity of 3.7
$L_\odot$. Possibly, L1489~IRS is a close binary, inhibiting the
formation of a circumstellar accretion disk
\citep*{jensen:binaries}. \citet{padgett:nicmos} only detect a single
point source, but with 0{\farcs}15 resolution, binaries with
separations less than 20~AU are still unresolved.

One possibility is that we observe L1489~IRS at the moment where it is
making its transition from an embedded object to a T~Tauri star, as
appears to be the case for HL~Tau. Rotating disks around T~Tauri stars
have typical radii of 500--1000 AU radius in line emission
\citep{handa:dmtau, mannings:haebe, dutrey:gmaur, guilloteau:dmtau},
but appear more compact ($\lesssim 100$ AU) when observed in continuum
due to differences in line and continuum emissivities. No sensitive
SCUBA images of the dust continuum around such T~Tauri stars have been
published, however, which may show more extended low-level structures,
and which offer a better comparison to the observations of
L1489~IRS. At a contraction velocity of only 0.5 km~s$^{-1}$, it will
take L1489~IRS's disk-like structure $2\times 10^4$ yr to contract to
500 AU. This is short compared to the $\sim 10^5$ yr typically
required to disperse the envelope, and only a detailed analysis of its
velocity structure will allow its detection in the presence of the
much larger (2--3 km~s$^{-1}$) rotation velocities.

Another possibility which such an analysis might confirm, is that
L1489~IRS's disk-like structure is not contracting but is fully
rotationally supported. If its central object is indeed a close binary
which has inhibited the formation of an accretion disk, this system
may find itself without an efficient means to carry away excess
angular momentum. Bipolar outflows driven by disk accretion are
generally thought to play an important role in this process
\citep[e.g.][]{shu:ppiii}. L1489~IRS's outflow as detected in CO 3--2
is very weak indeed \citep{mrh:taurus2}. A binary companion may have
cleared a gap in the accretion disk, shutting off or greatly
decreasing the inward flow of material. Ascertaining whether L1489~IRS
is, as this would suggest, a young analog to the GG~Tau `ring' system
\citep*{guilloteau:ringworld} will require additional observations in
the near- and mid-infrared characterizing this object's multiplicity
and accretion state, as well as aperture-synthesis in molecular lines
to probe the velocities in the disk-like structure.

\subsection{The inner regions of the envelopes\label{s:central}}

Section \ref{s:models} only considered models which describe the
emission over the entire envelope in a spherically symmetric geometry,
and neglected any radial changes in dust properties or source
geometry. This section adopts a different view of the data, fitting
two Gaussians to the data, one describing the emission from the
central region and one from the extended emission. By minimizing the
size of the first component, we maximize the difference between the
emission from the central region and the extended emission, and thus
the difference with the models explored in the previous two sections.
This description of the intensity distribution only refers to the
central $\sim 40''$ around the objects. Since the emission follows a
radial power law (\S \ref{s:models}, a series of Gaussians would be
required to fit the entire images. By limiting ourselves to the inner
$40''$, two Gaussians suffice. We do not claim that these Gaussian
carry physical significance beyond being a description of the emission
profiles and allowing for an investigation of the dust properties on
small scales.

Table \ref{t:flux} lists the (deconvolved) major and minor axes and
position angles of the central Gaussian components; the listed
components fit the radial emission profiles in the inner $40''$ as
well as the single power laws of \S \ref{s:powmdl}. We consider a
simple isothermal model for the central regions, with an effective
temperature estimated from the infrared SED. Warm material near the
star dominates the emission at 100 $\mu$m and shorter, yielding a
characteristic temperature for the inner regions of the envelope.
This is essentially an effective temperature: it describes the
characteristics of the emission, not the physical temperature of the
bulk of the material.  Using an effective temperature, allows us to
link the observed intensity in both wavelength bands through a
spectral index of the dust emissivity.  The free parameters of this
model are the effective temperature, the source diameter, and the
spectral index of the dust emissivity, again parameterized as $\kappa
= 0.1 (\nu/10^{12}\,{\rm Hz})^\beta$ cm$^{2}$~g$^{-1}$ (gas+dust, with
gas:dust=100). We assume that the emission is optically thin (but see
below).

The 850 and 450 $\mu$m fluxes and sizes of the central peaks and the
infrared SED from \citet{kenyon:sed_1} yields best-fit results as
listed in Table \ref{t:central}.  The obtained diameters are $\sim
7''$ or 1000 AU, slightly larger than the typical value for disks
around T~Tauri stars as seen in molecular lines \citep{handa:dmtau,
mannings:haebe, dutrey:gmaur, guilloteau:dmtau}. The spectral index of
the dust emissivity $\beta$ is 0.9--1.1, with an estimated uncertainty
of 0.2. Toward L1527~IRS the low $\beta$ is directly reflected in the
spectral index maps of Fig. \ref{f:maps}; toward the other sources
increased temperature partially compensates for the lower
$\beta$. These values of the dust emissivity's spectral index are
significantly smaller than the 1.5--1.9 representative of the entire
envelope (Tables \ref{t:powfit} and \ref{t:shufit}), and indicate that
grain growth may take place on small scales.

Alternatively, the low spectral index indicates a significant
contribution to the emission from optically thick material. The
envelope models as derived in the previous sections are optically thin
at 850 and 450 $\mu$m, even at small radii, and significant flattening
is required to increase the opacity without changing the flux. Part of
the central flux could also originate in an unresolved, optically
thick circumstellar disk. Previous 3~mm interferometer observations
\citep{terebey:disks, ohashi:taurusnma96, mrh:taurus1} and limits
inferred by \citet{chandler:scuba} suggest, however, that no more than
10--20\% of the flux at 850 and 450 $\mu$m can be attributed to an
unresolved point source. We therefore conclude that the observed
decrease in spectral index on small scales is likely caused by a
change in dust properties, but cannot rule out a significant
contribution from optically thick material if the inner regions of the
envelope are highly flattened.

\subsection{The nature of L1489~NE-SMM and L1535~N-SMM\label{s:coreb}}

The presence of L1489~NE-SMM and L1535~N-SMM within 10,000 AU
of the YSOs L1489~IRS and L1535~IRS, respectively, shows that star
formation in truly isolated cores is rare even in Taurus. These cores
are not chance alignments but are physically close to the
YSOs. Moderately dense ($\sim 10^5$ cm$^{-3}$) gas traced by HCO$^+$
1--0 shows emission from these cores near the systemic velocities of
the YSOs \citep{mrh:taurus1}, and L1535~N-SMM is visible in
scattered near-infrared emission in observations by
\citet{hodapp:kprime}. Their presence was already noted by
\citet{ladd:firsubmm2}, who show that they coincide with NH$_3$ cores
detected by \citet{benson:densecores} and \citet*{ungerechts:nh3}.

The nature of these cores is best investigated through the residual
images after subtracting the YSO models (Fig. \ref{f:shures}). Because
the power-law model and the inside-out collapse model provide equally
well fits to the continuum data, only the residuals of the latter are
shown. Figure \ref{f:2azim} presents the radial flux density profiles
around the centers of these cores. Contrary to the emission
distribution of the YSOs, those of L1489~NE-SMM and
L1535~N-SMM are not characterized by a power-law, but more closely
resemble a Gaussian of deconvolved FWHM $180''\times 70''$ and
$200''\times 100''$, respectively. It seems unlikely that this shape
is caused by the assumption of a Gaussian distribution in the model
fits, since these only applied to the $80''\times 80''$ immediately
around the YSOs.

The spectral index of the emission from these cores is not
significantly changed by the subtraction of the best fit YSO models,
and indicates a dust temperature of 10--15 K, assuming a dust
emissivity spectral index of $\beta =1$--2. For such low temperatures,
masses of 0.5--3.0 $M_\odot$ are found for both cores, significantly
larger than the circumstellar masses of L1535~IRS and L1489~IRS, and
close to the virial mass estimates assuming a velocity dispersion of
0.2--0.3 km~s$^{-1}$ and a density distribution given by $1/r$.  The
low temperature throughout the cores, their large masses, and the lack
of central concentration all suggest that these cores do not contain
embedded sources. Further interferometric searches for embedded
continuum sources and molecular line observations of possible infall
signatures will shed more light on the nature of these cores.

Contrary to L1489~IRS and L1535~IRS, the fields of L1527~IRS and TMC~1
are dominated by the YSO's themselves. We interpret the residual
emission toward L1527~IRS as dust heated to $\gtrsim 30$ K along the
edge of the outflow. An upper limit to the mass of 0.2 $M_\odot$ is
found, or 10\% of the total inferred envelope mass. This is comparable
to the total amount of swept up CO \citep[0.2
$M_\odot$;][]{mrh:taurus2}, and may be the same material.  No residual
emission is detected toward TMC~1, making it an ideal target for
further studies of isolated star formation.

\section{Conclusions\label{s:conclusions}}

This paper presents high-quality submillimeter-continuum observations
of four embedded YSOs in Taurus, analyzes them in terms of the popular
collapse model of \citet{shu:selfsim}, and compares the results to
molecular-line data. The main conclusions can be summarized as
follows.

\begin{enumerate}
\item The emission around the YSOs can be described by $\sim 1000$ AU
elongated central peaks, oriented perpendicular to the outflows,
embedded in extended emission filling the fields of view. The density
distribution in the extended envelopes is characterized by a radial
power-law the with negative index $p=1$--2, and outer radii between
2000 AU and 12,000 AU. The inferred envelope masses are 0.016--3.5
$M_\odot$.

\item Equally good fits can be found using the inside-out collapse
model of \citet{shu:selfsim}, yielding sound speeds $a=0.19$--0.46
km~s$^{-1}$ and ages between $3\times 10^4$ and $2\times 10^6$ yr.

\item Only independent measurements of the density and the velocity
structure can offer true tests of collapse models. Combined with
previously obtained molecular line data, the SCUBA observations
provide such a test. The combination of both data sets confirms that
L1527~IRS, TMC~1, and probably L1535~IRS, are surrounded by collapsing
envelopes which can be described by the inside-out collapse model.

\item For one source, L1489~IRS, the inside-out collapse model
spectacularly fails to fit the continuum and line data simultaneously.
Instead, a better solution is offered by a rotating, disk-like
structure of 2000 AU radius with a density distribution following a
radial power-law. We suggest that this source may be in a transitional
phase between a Class~I and Class~II object, and is possibly stunted
in its evolution because of insufficient means to carry away excess
angular momentum.

\item Grain growth on small scales is evidenced by the decrease of the
spectral index of the dust emissivity from 1.5--2.0 in the extended
envelopes to $1.0\pm 0.2$ in the central elongated cores, although
high optical depth in the inner regions of the envelopes is also a
possible explanation.

\item Star formation in isolated cores is rare, even in Taurus, as is
witnessed by the presence of two, apparently starless, cores within
10,000 AU of our YSOs. These cores are cold (10--15 K), lack central
concentration in their density structure, but have sufficient mass to
eventually form stars (0.5--3.0 $M_\odot$).

\end{enumerate}

\acknowledgments The authors gratefully acknowledge the staff of the
JCMT for carrying out the SCUBA observations. The research of MRH is
supported by the Miller Institute for Basic Research in Science. Ewine
van Dishoeck is thanked for a careful reading of the manuscript and
many valuable comments. Yancy Shirley is thanked for useful
discussions. The referee, Neal Evans, provided valuable comments which
greatly improved the paper.

\newpage


\figcaption[f_maps.ps]{Images of the $\lambda=850$ $\mu$m and 450
$\mu$m emission observed with SCUBA on the JCMT toward the YSOs
L1489~IRS, L1535~IRS, L1527~IRS, and TMC~1. The contours are 2$\sigma$
intervals as listed in Table \ref{t:flux}. The right-hand panels show
the spectral index between 450 $\mu$m and 850 $\mu$m as derived from
the data, plotted with contour intervals of 0.5. Note that the center
of the images of L1489~IRS is $(30'',40'')$ to better bring out the
extended emission. The filled circles in the lower right of the panels
indicate the main beam size.\label{f:maps}}

\figcaption[f_central.ps]{Blow-up of the central $40''\times 40''$ of
the 450 $\mu$m images. Contour levels are drawn at the same 2$\sigma$
intervals as Fig. \ref{f:maps}. The vertical bars indicate the FWHM
sizes of the beams; the arrows indicate the direction of the
outflows.\label{f:central}}

\figcaption[f_azim.ps]{({\it left\/}) Radial profiles of the 850 and
450 $\mu$m emission obtained after azimuthal averaging of the SCUBA
data around the source centers. For L1527~IRS this only includes a
one-beam wide strip perpendicular to the outflow to minimize
contamination with emission from emission along outflow. The solid
symbols show the data, binned in half beam-width intervals. The error
bars show the standard deviation within each annulus, which includes
noise and any deviations from circular symmetry in the source. The
dotted curve corresponds to the simple fit to the JCMT beam as derived
from maps of Uranus (Table \ref{t:beam}). The dash-dotted and solid
lines depict the best-fit results for a power-law model (\S
\ref{s:powmdl}) and the inside-out collapse model (\S \ref{s:shumdl}),
respectively. The emission peaks at $\log(r)\approx 1.9$ toward
L1489~IRS and L1535~IRS correspond to the adjacent cores L1489~NE-SMM
and L1535~N-SMM. ({\it right\/}) Radial profiles of the 850--450
$\mu$m spectral index, with error bars including noise and deviations
from circular symmetry. The calibration uncertainty between both
wavelength bands in not included and may shift the spectral index up
or down by 0.4, but does not affect its radial
distribution. \label{f:azim}}

\figcaption[f_spec.ps]{Observed single-dish spectra from
\citet{mrh:taurus1,mrh:taurus2} ({\it histograms\/}) and line profiles
predicted by the inside-out collapse model including CO depletion
({\it solid curves\/}). Good agreement between the observed spectra
and the model predictions is found for L1527~IRS (a) and TMC~1 (b),
when CO is depleted by a factor of 10 in the coldest regions. Because
of L1535~IRS's weak continuum emission compared to L1535~IRS~B, the
derived model parameters are uncertain, and the predicted line
profiles differ from the observations (c). For L1489~IRS (d) the
predicted profiles do not agree with the observations. Instead, a
different model is proposed for this source in the
text. \label{f:spec}}

\figcaption[f_shures.ps]{Residual images after subtracting the
best-fit inside-out collapse models of \S \ref{s:shumdl}. For
L1489~IRS and L1535~IRS, the Gaussian background source used in the
fitting procedure has not been subtracted out.  The contours are the
same 2$\sigma$ intervals as Fig. \ref{f:maps} and as listed in
Table\ref{t:flux}.\label{f:shures}}

\figcaption[f_2azim.ps]{Radial profiles of L1489~NE-SMM and
L1535~N-SMM, after subtraction of the best-fit power-law ({\it
symbols\/}) and inside-out collapse ({\it solid line\/}) models. The
dotted line shows the beam profile.\label{f:2azim}}

\newpage


%
%
%
%

\begin{deluxetable}{lrrlrr}
\tablewidth{0pt}
\tablecolumns{8}
\tablecaption{Observations\label{t:obs}}
\tablehead{
\colhead{Source} & \colhead{$\alpha$(2000.0)} & \colhead{$\delta$(2000.0)} & 
\colhead{Dates} & \colhead{$N_{\rm int}$} & \colhead{Chop PA}
}
\startdata
L1489~IRS & ${\rm 04^h 04^m 43{\fs}10}$ & $+26^\circ 18' 56{\farcs}9$ & 
  1997/Dec/7,8,10,16 & 60 & $60^\circ$, $45^\circ$ \\
L1535~IRS & 04 35 35.23 & +24 08 25.3 & 
  1998/Jan/18 & 40 & $0^\circ$ \\
L1527~IRS & 04 39 53.89 & +26 03 11.0 &
  1997/Dec/5,6 & 15 & AZ \\
TMC~1 & 04 41 12.74 & +25 46 33.3 &
  1997/Dec/7,8,16,18 & 35 & AZ \\
\enddata
\end{deluxetable}

%
%
%
%

\begin{deluxetable}{lrc}
\tablewidth{0pt}
\tablecolumns{3}
\tablecaption{Simple description of beam pattern\label{t:beam}}
\tablehead{
\colhead{$\lambda$} & \colhead{Relative} & \colhead{FWHM} \\
\colhead{($\mu$m)} & \colhead{amplitude} & \colhead{($''$)}}
\startdata
850 & 0.970 & 14.5 \\
    & 0.020 & 40 \\
    & 0.010 & 70 \\
450 & 0.934 & 8 \\
    & 0.060 & 30 \\
    & 0.006 & 120 \\
\enddata
\end{deluxetable}

%
%
%
%

\begin{deluxetable}{lrrrrrcr}
\tablewidth{0pt}
\tablecolumns{8}
\tablecaption{Noise, flux, and Gaussian fits to emission peak\label{t:flux}}
\tablehead{
 & \colhead{$\lambda$} & \colhead{noise rms} &
\colhead{$F_{\rm total}$\tablenotemark{a}} & 
\colhead{$F_{\rm peak}$\tablenotemark{b}} & 
\colhead{$\int F d\Omega$\tablenotemark{b}} &
\colhead{Size\tablenotemark{b}} & 
\colhead{PA\tablenotemark{b}} 
\\
\colhead{Source} & \colhead{($\mu$m)} & \colhead{(Jy~beam$^{-1}$)} &
\colhead{(Jy)} & 
\colhead{(Jy~beam$^{-1}$)} & \colhead{(Jy)} & \colhead{($''$)} &
\colhead{($^\circ$)}
}
\startdata
L1489~IRS & 850 & 0.02 &  5.78 & 0.36 & 0.43 & $ 8\times  6$ & 54 \\
          & 450 & 0.10 & 23.3  & 1.76 & 2.82 & $ 8\times  4$ & 51 \\
L1535~IRS & 850 & 0.03 &  6.03 & 0.18 & 0.29 & $12\times  9$ & 53 \\
          & 450 & 0.15 & 22.4  & 0.36 & 1.00 & $12\times  9$ & 38 \\
L1527~IRS & 850 & 0.06 & 10.6  & 0.84 & 1.04 & $ 9\times  5$ & 24 \\
          & 450 & 0.27 & 65.3  & 2.79 & 5.07 & $10\times  5$ & 36 \\
TMC~1     & 850 & 0.02 &  2.60 & 0.19 & 0.24 & $13\times 11$ & 40 \\
          & 450 & 0.10 & 12.5  & 0.65 & 1.20 & $10\times  5$ & 56 \\
\enddata
\tablenotetext{a}{Total flux over entire observed image.}
\tablenotetext{b}{Parameters of smallest Gaussian which can be fit 
to central peak. Size is deconvolved FWHM. The estimated error in the fitted
peak fluxes are 5\%, in the FWHM at 0{\farcs}3--0{\farcs}5, and in the PA
are $5^\circ$.}
\end{deluxetable}

%
%
%
%

\begin{deluxetable}{lrrrrrrr}
\tablewidth{0pt}
\tablecolumns{8}
\tablecaption{Best-fit parameters power-law models\label{t:powfit}}
\tablehead{
 & \colhead{$\log n_0$} & & \colhead{$M_{\rm env}$} & &
\colhead{$F(850)$\tablenotemark{a}} & & \colhead{FWHM\tablenotemark{a}}
\\ 
\colhead{Source} & \colhead{(cm$^{-3}$)} & \colhead{$p$} & 
\colhead{($M_\odot$)} & \colhead{$\beta$} & \colhead{(Jy)} & 
\colhead{$\alpha$\tablenotemark{a}} & \colhead{($''$)}}
\startdata
L1489~IRS & 
  $5.0 \pm 0.1$ & $2.1 \pm 0.2$ & $0.016 \pm 0.004$ & 
  $1.5 \pm 0.2$ & $0.22 \pm 0.02$ & $1.9 \pm 0.2$ & $162\times 88 \pm 20$\\
L1535~IRS & 
  $4.8 \pm 0.1$ & $1.7 \pm 0.2$ & $0.064\pm 0.003$ &
  $1.1 \pm 0.2$ & $0.18 \pm 0.02$ & $2.3 \pm 0.2$ & $180\times 148 \pm 20$\\
L1527~IRS & 
  $6.2\pm 0.1$ & $0.9\pm 0.2$ & $4.7\pm 1.0$ & $2.0\pm 0.3$ &
  n/a & n/a & n/a \\
TMC~1 & 
  $5.1\pm 0.1$ & $1.2\pm 0.1$ & $0.58\pm 0.1$ & $1.5\pm 0.2$ &
  n/a & n/a & n/a \\
\enddata
\tablenotetext{a}{Gaussian model for adjacent core.}
\tablecomments{Using $R_{\rm out}=2000$ AU (L1489~IRS), 8000 AU 
  (L1535~IRS, L1527~IRS), 12,000 AU (TMC~1).}
\end{deluxetable}

%
%
%
%

\begin{deluxetable}{lrrrrrrr}
\tablewidth{0pt}
\tablecolumns{8}
\tablecaption{Best-fit parameters collapse models\label{t:shufit}}
\tablehead{
 & \colhead{$a$} & \colhead{$\log (r_{\rm CEW})$} &
\colhead{$\log (t)$} & \colhead{$M_{\rm env}$} & & 
  \colhead{$F(850)$\tablenotemark{a}} 
\\
\colhead{Source} & \colhead{(km~s$^{-1}$)} & \colhead{(AU)} &
 \colhead{(yr)} & \colhead{($M_\odot$)} & \colhead{$\beta$} & \colhead{(Jy)} & 
\colhead{$\alpha$\tablenotemark{a}} }
\startdata
L1489~IRS & 
  $0.46\pm 0.04$ & $5.3 \pm 0.2$ & $6.3\pm 0.2$ & $0.023\pm 0.006$ &
  $1.4\pm 0.2$ & $0.23 \pm 0.03$ & $1.9\pm 0.2$ \\
L1535~IRS & 
  $0.29\pm 0.04$ & $4.6\pm 0.1$ & $5.8\pm 0.2$ & $0.19\pm 0.06$ &
  $1.7\pm 0.2$ & $0.18\pm 0.02$ & $2.0\pm 0.2$ \\
L1527~IRS & 
  $0.47\pm 0.04$ & $3.6\pm 0.2$ & $4.6\pm 0.1$ & $3.0\pm 0.5$ &
  $1.8\pm 0.2$ & n/a & n/a \\
TMC~1 & 
  $0.19 \pm 0.03$ & $4.0\pm 0.1$ & $5.4\pm 0.2$ & $0.59\pm 0.2$ &
  $1.5\pm 0.2$ & n/a & n/a \\
\enddata
\tablenotetext{a}{Gaussian model for adjacent core. FWHM unchanged from
   values in Table \ref{t:powfit}}
\tablecomments{Using $R_{\rm out}=2000$ AU (L1489~IRS), 8000 AU 
  (L1535~IRS, L1527~IRS), 12,000 AU (TMC~1).}
\end{deluxetable}

\begin{deluxetable}{lrrr}
\tablewidth{0pt}
\tablecolumns{4}
\tablecaption{Parameters obtained previously from fits to line profiles 
only\label{t:oldshu}}
\tablehead{
 & \colhead{a} & \colhead{$\log(r_{\rm CEW})$} & \colhead{$\log(t)$} \\
\colhead{Source} & \colhead{(km~s$^{-1}$)} & \colhead{(AU)} &
\colhead{(yr)}}
\startdata
L1489~IRS & 0.23 & 4.2 & 5.5 \\
L1535~IRS & \nodata & \nodata & \nodata \\
L1527~IRS & 0.23 & 3.9 & 5.2 \\
TMC~1 & 0.17 & 4.0 & 5.4 \\
\enddata
\tablerefs{All parameters from \citealt{mrh:thesis}}
\end{deluxetable}

%
%
%
%

\begin{deluxetable}{lrrrrr}
\tablewidth{0pt}
\tablecolumns{6}
\tablecaption{Model fits to central peak\label{t:central}}
\tablehead{
 & \colhead{Diameter} & \colhead{$T_{\rm dust}$} & & 
\colhead{$L_{\rm bol}$} & \colhead{$M_{\rm central}$} \\
\colhead{Source} & \colhead{($''$)} & \colhead{(K)} & \colhead{$\beta$} &
\colhead{($L_\odot$)} & \colhead{($M_\odot$)}}
\startdata
L1489~IRS &  5.5 & 45 & 1.1 & 1.75 & 0.01 \\
L1535~IRS &   11 & 41 & 0.9 & 0.51 & 0.006 \\
L1527~IRS &  6.5 & 35 & 0.9 & 0.95 & 0.035 \\
TMC~1     &  7.0 & 36 & 0.8 & 0.23 & 0.006 \\
\enddata
\end{deluxetable}


\input psfig

\newpage

\begin{figure}
\figurenum{\ref{f:maps}}
\begin{center}
\leavevmode
\psfig{figure=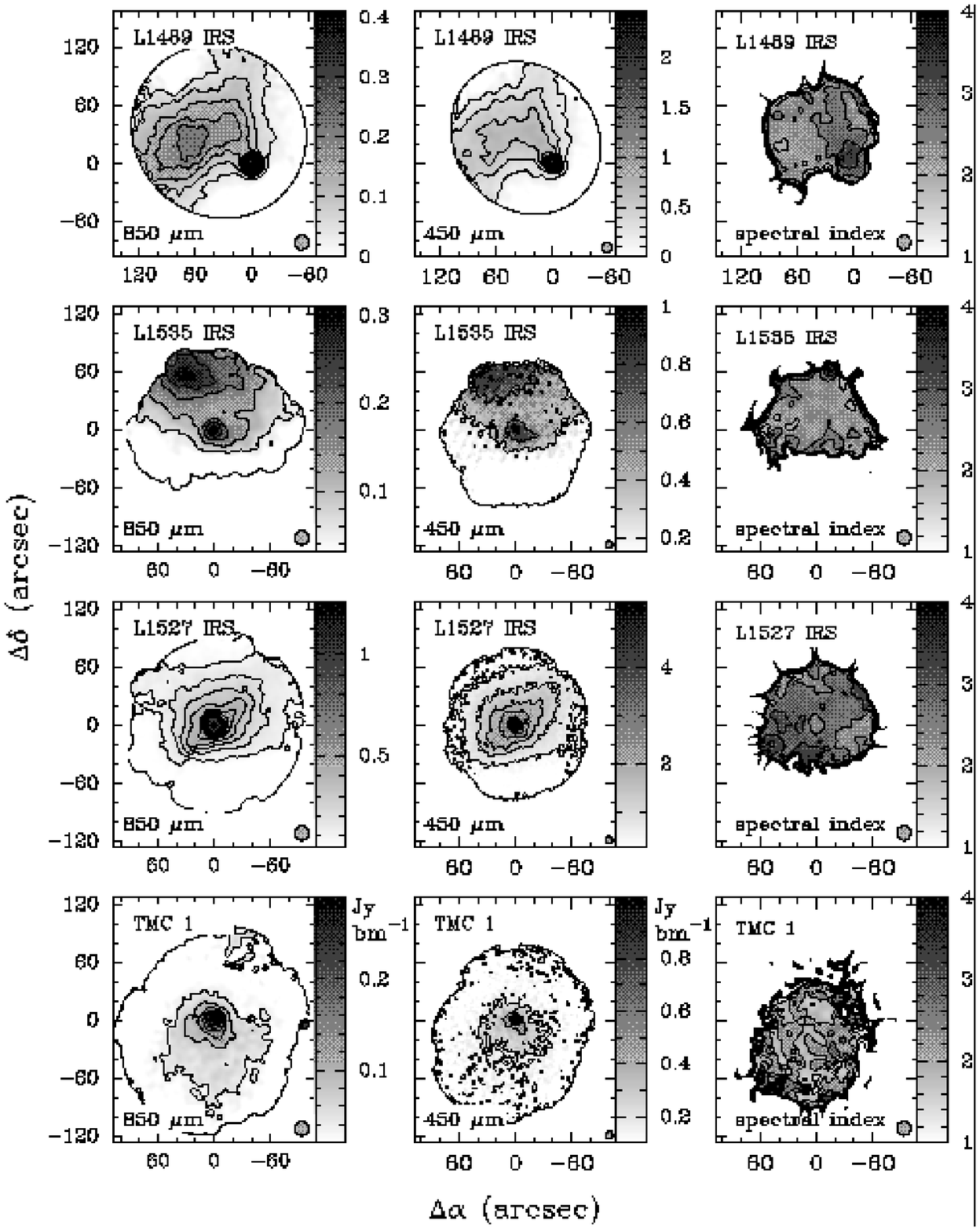,height=18truecm,angle=0}
\end{center}
\caption{}
\end{figure}

\begin{figure}
\figurenum{\ref{f:central}}
\begin{center}
\leavevmode
\psfig{figure=fig2.ps,height=10truecm,angle=0}
\end{center}
\caption{}
\end{figure}

\begin{figure}
\figurenum{\ref{f:azim}}
\begin{center}
\leavevmode
\psfig{figure=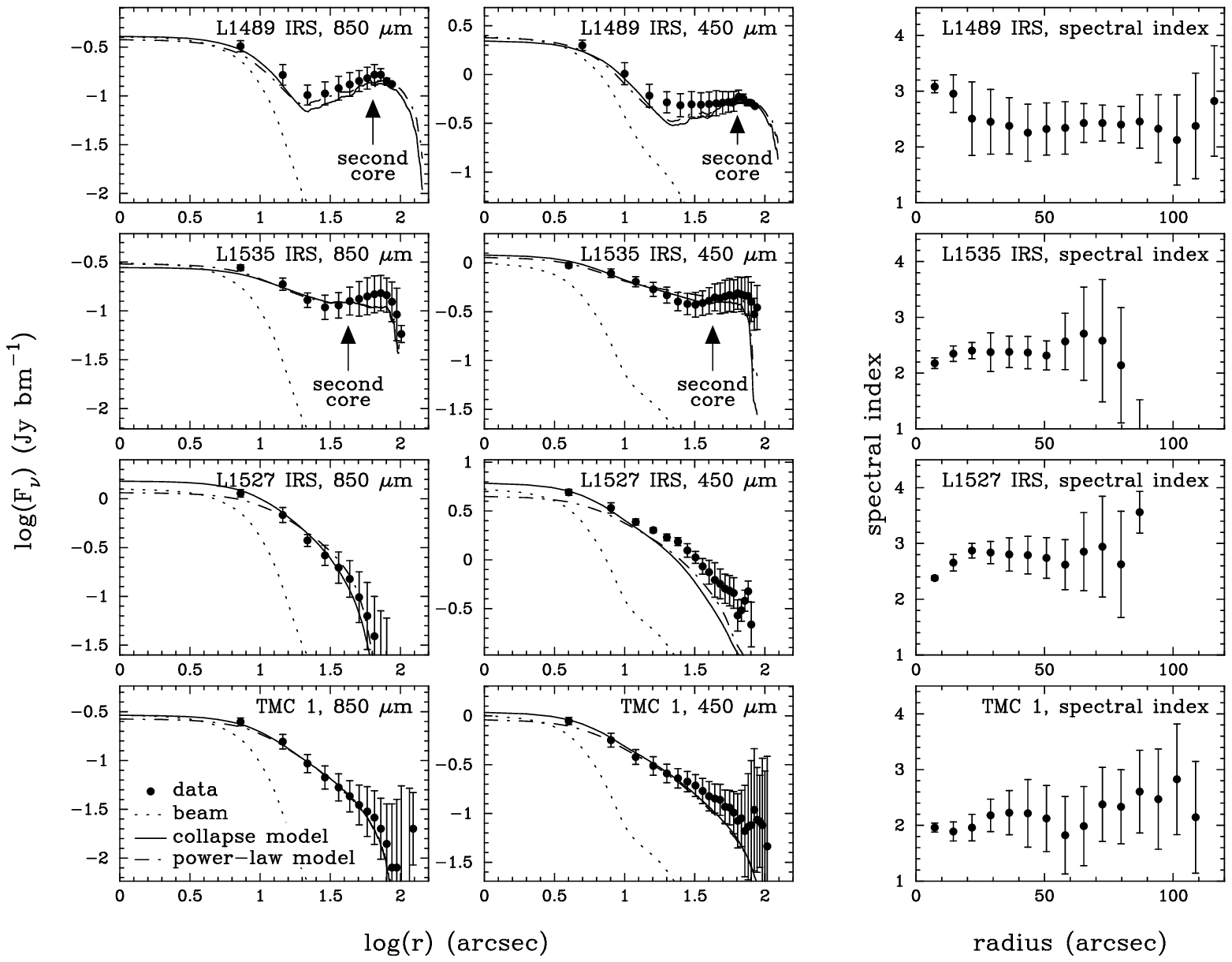,height=16truecm,angle=0}
\end{center}
\caption{}
\end{figure}

\begin{figure}
\figurenum{\ref{f:spec}a}
\begin{center}
\leavevmode
\psfig{figure=fig4a.ps,height=18truecm,angle=0}
\end{center}
\caption{}
\end{figure}

\begin{figure}
\figurenum{\ref{f:spec}b}
\begin{center}
\leavevmode
\psfig{figure=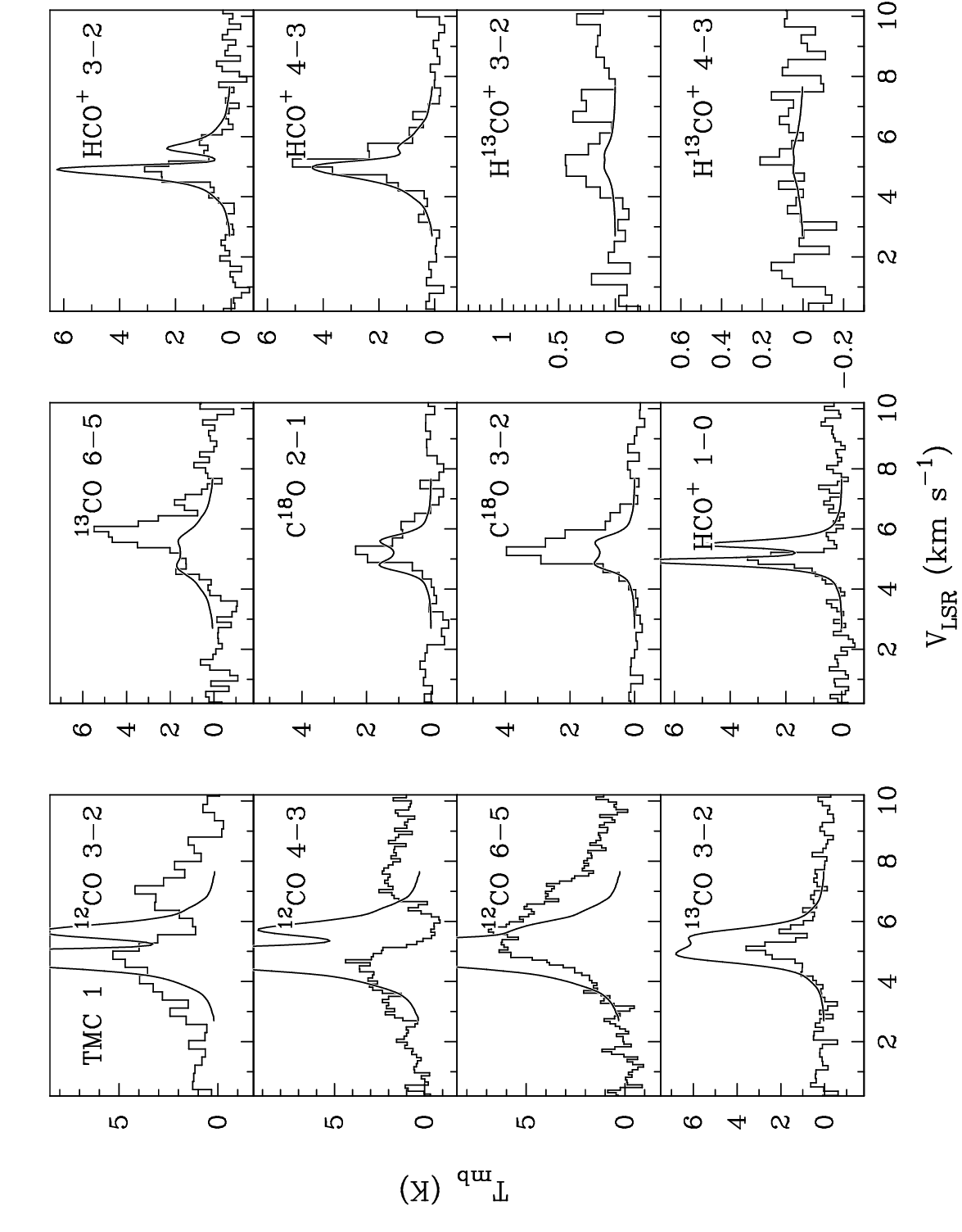,height=18truecm,angle=0}
\end{center}
\caption{}
\end{figure}

\begin{figure}
\figurenum{\ref{f:spec}c}
\begin{center}
\leavevmode
\psfig{figure=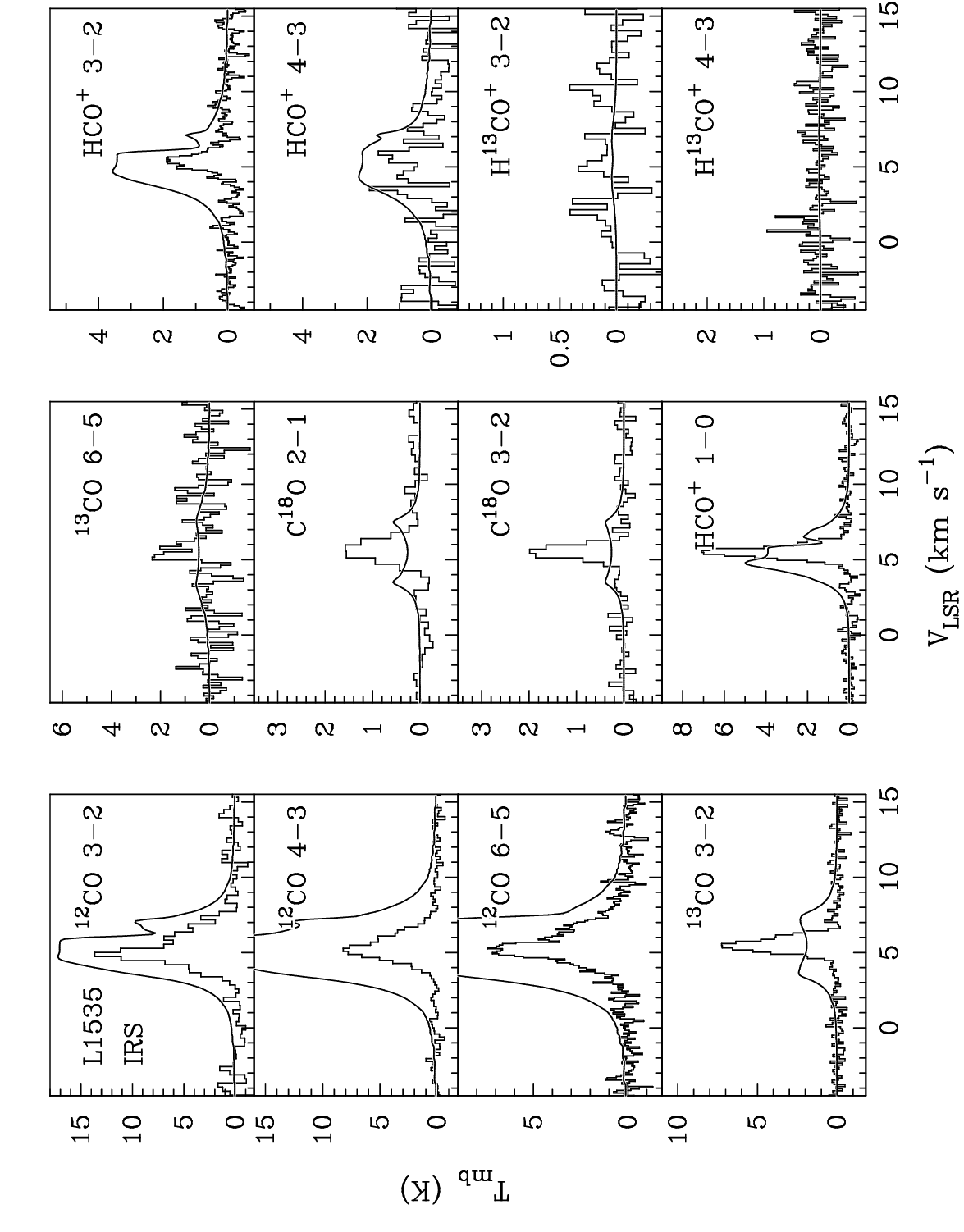,height=18truecm,angle=0}
\end{center}
\caption{}
\end{figure}

\begin{figure}
\figurenum{\ref{f:spec}d}
\begin{center}
\leavevmode
\psfig{figure=fig4d.ps,height=18truecm,angle=0}
\end{center}
\caption{}
\end{figure}

\begin{figure}
\figurenum{\ref{f:shures}}
\begin{center}
\leavevmode
\psfig{figure=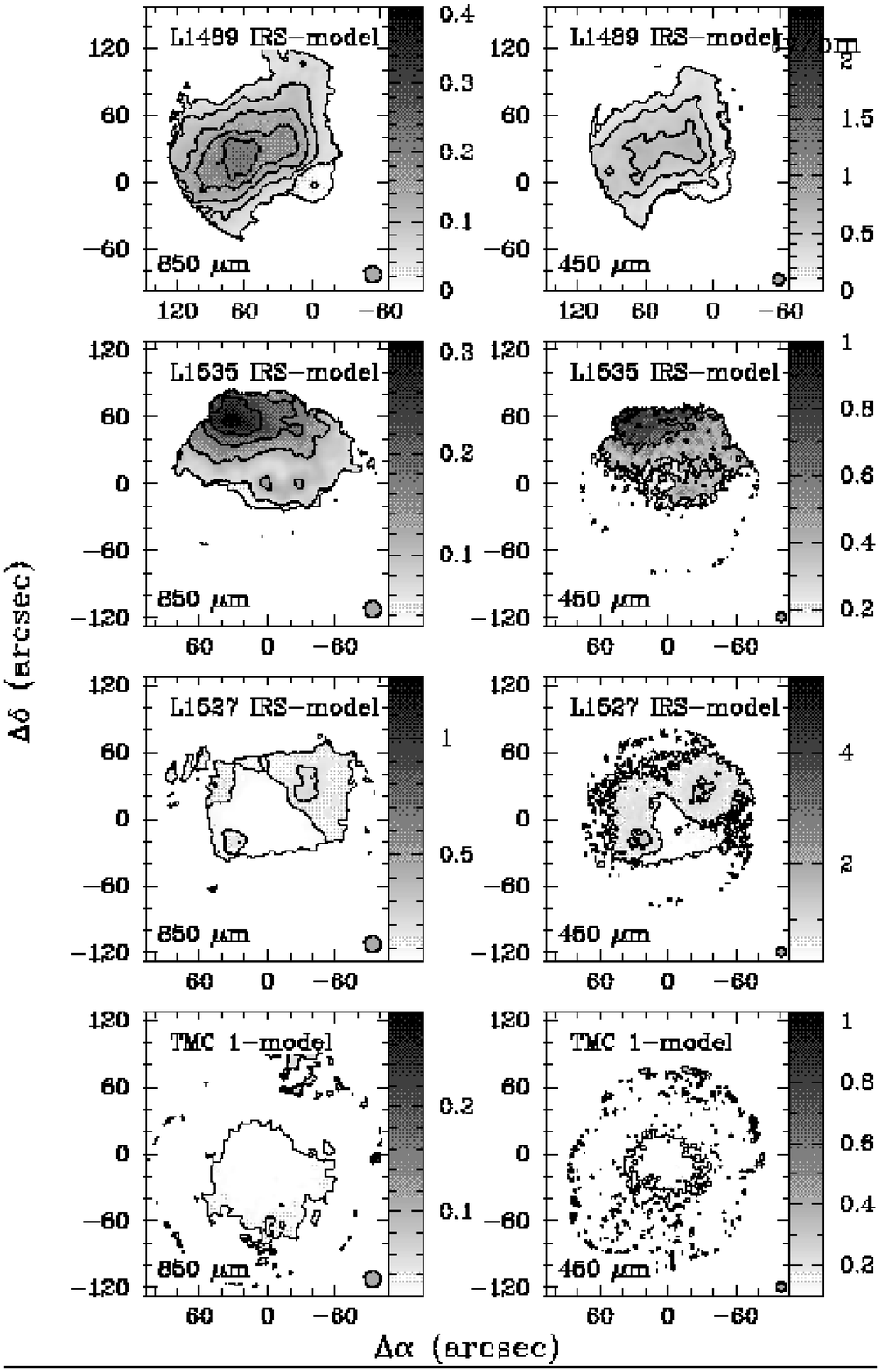,height=18truecm,angle=0}
\end{center}
\caption{}
\end{figure}

\begin{figure}
\figurenum{\ref{f:2azim}}
\begin{center}
\leavevmode
\psfig{figure=fig6.ps,height=18truecm,angle=0}
\end{center}
\caption{}
\end{figure}

\end{document}